\documentclass[preprint,11pt]{elsarticle}

\usepackage[table,xcdraw]{xcolor}
\usepackage{graphicx}
\usepackage{lscape}
\usepackage{upgreek}
\usepackage{caption} 
\usepackage{hyperref}
\usepackage{todonotes}
\usepackage{amssymb}
\usepackage{lineno}
\usepackage{url}
\usepackage{array}
\usepackage{enumitem}
\usepackage{multirow}
\usepackage{soul}
\usepackage{float}
\usepackage{color}
\usepackage{pgfplots}
\usepackage{booktabs}
\usepackage[export]{adjustbox}
\usepackage{graphicx}
\usepackage{colortbl}
\usepackage{array}
\usepackage{booktabs}
\usepackage{comment}
\usepackage{multirow}
\usepackage{xurl}
\usepackage{soul}

\usepackage[normalem]{ulem}

\definecolor{pblue}{rgb}{0.13,0.13,1}
\definecolor{pgreen}{rgb}{0,0.5,0}
\definecolor{pred}{rgb}{0.9,0,0}
\definecolor{pgrey}{rgb}{0.46,0.45,0.48}
\definecolor{codebackground}{rgb}{0.95, 0.95, 0.92}
\definecolor{gray50}{gray}{.5}
\definecolor{gray40}{gray}{.6}
\definecolor{gray30}{gray}{.7}
\definecolor{gray20}{gray}{.8}
\definecolor{gray10}{gray}{.9}
\definecolor{gray05}{gray}{.95}
\definecolor{arsenic}{rgb}{0.23, 0.27, 0.29}

\definecolor{darkmagenta}{rgb}{0.55, 0.0, 0.55}
\definecolor{darkgreen}{RGB}{6, 46, 3}
\definecolor{amber}{rgb}{1.0, 0.75, 0.0}

\usepackage{hyperref}
\hypersetup{
pdfproducer={},  
}

\usepackage{tikz}
\definecolor{darkpastelpurple}{rgb}{0.59, 0.44, 0.84}
\tikzstyle{mybox} = [draw=black, very thick, rectangle, rounded corners, inner ysep=5pt, inner xsep=5pt]
\usepackage{color}

\journal{Journal of Systems and Software}

\newcommand{\quickwordcount}{%
  \immediate\write18{texcount -1 -sum -merge \jobname.tex > \jobname-words.sum }%
  \input{\jobname-words.sum} words%
}

\begin{document}
\begin{frontmatter}

\markboth{Moreschini et al.}{: A Systematic Mapping Study}

\title{Edge to Cloud Tools:\\A Multivocal Literature Review}

\author{Sergio Moreschini,$^{1*}$, Elham Younesian,$^{1,2*}$ David Hästbacka,$^1$ \\ Michele Albano,$^3$  Jiří Hošek,$^2$ Davide Taibi$^{1,4}$}

\address{$^1$Tampere University, Tampere (Finland) --- $^{2}$Brno University of Technology, Brno (Czech Republic)\\
$^{3}$Aalborg University, Aalborg (Denmark) --- $^{4}$ University of Oulu, Oulu (Finland) \\
sergio.moreschini@tuni.fi, elham.younesian@vut.cz, david.hastbacka@tuni.fi, mialb@cs.aau.dk, hosek@feec.vutbr.cz ,davide.taibi@oulu.fi\\
$*$ the two authors equally contributed to the paper}
	
\begin{abstract}

Edge-to-cloud computing is an emerging paradigm for distributing computational tasks between edge devices and cloud resources. Different approaches for orchestration, offloading, and many more purposes have been introduced in research. However, it is still not clear what has been implemented in the industry. This work aims to merge this gap by mapping the existing knowledge on edge-to-cloud tools by providing an overview of the current state of research in this area and identifying research gaps and challenges. For this purpose, we conducted a Multivocal Literature Review (MLR) by analyzing 40 tools from 1073 primary studies (220 PS from the white literature and 853 PS from the gray literature). We categorized the tools based on their characteristics and targeted environments. Overall, this systematic mapping study provides a comprehensive overview of edge-to-cloud tools and highlights several opportunities for researchers and practitioners for future research in this area.

\end{abstract}


\end{frontmatter}

\section{Introduction}
\label{sec:Introduction}
Edge computing is an emerging computing paradigm where data processing and storage are performed closer to the source rather than on centralized servers in the cloud. This is achieved by placing computing resources at the network edge, which enables the processing of data to be performed in real-time and with low latency. 

Edge-to-cloud tools refer to the combination of hardware, software, and services used to collect, process, and analyze data from edge devices (sensors, machines, etc.) and transmit that data to cloud platforms for further processing and analysis. Edge-to-cloud tools have emerged as a key technological solution for modern-day data processing and analysis needs. The combination of edge devices, cloud platforms, and software services has enabled the collection, processing, and analysis of data from distributed edge devices in real-time. The characteristics of edge-to-cloud tools vary depending on the specific use case and organization's needs. These characteristics can include processing and analyzing data at the edge of the network, reducing latency, optimizing data transmission, scalability, and flexibility. Additionally, edge-to-cloud tools can provide access to advanced analytics and machine learning algorithms, enabling organizations to gain insights and make data-driven decisions. In this context, understanding the different characteristics of edge-to-cloud tools such as offloading, orchestration, workflow management and other computational tasks in the cloud continuum is crucial to harness their full potential for various applications.



In order to solve the aforementioned issues, we performed a systematic mapping study from 1073 primary studies in academic and industrial (i.e., gray literature) sources to classify the edge-to-cloud tools in the cognitive cloud continuum. Therefore, the goal of this work is to contribute to the state-of-the-art by providing a comparison of edge-to-cloud tools for both researchers and practitioners. First of all, such a comparison provides the reader with a list of all the available tools. Following this, by performing a comparison, the reader can understand the main characteristics of each tool, including its license. Another fundamental aspect that we aim to tackle is the target environment of each tool so that users can understand which architecture a tool is compatible with.

The rest of this paper is structured as follows. Section~\ref{sec:Background} introduces the background with a particular focus on concepts such as Cloud Continuum and Edge-to-Cloud Offloading. Section~\ref{sec:ResearchQuestions} presents the main goal of the work and the related research questions. Section~\ref{sec:StudyDesign} reports the method, and the multiple steps performed to follow it, used in this paper to answer the 3 research questions proposed. Section~\ref{sec:StudyResults}, illustrates the results obtained by answering the different research questions. The main discussion points, future challenges and possible threats to validity. The work ends with the conclusions presented in Section~\ref{sec:Conclusion}.
\section{Background}
\label{sec:Background}
In this Section, we introduce the background of this work in cloud, edge, cloud continuum, and edge-to-cloud technologies.

Cloud continuum is the seamless integration of different cloud services and resources, such as IoT devices, fog, and edge nodes. Edge and cloud technologies have grown significantly during these years. Therefore, investigating offloading between these environments has become important for practitioners and academics.
In this Section, we provide an overview of the cloud continuum and edge-to-cloud offloading.

\subsection{Cloud continuum}
Cloud Computing was defined officially in 2011 by the National Institute of Standard and Technologies (NIST) as \textit{"a model for enabling ubiquitous, convenient, on-demand network access to a shared pool of configurable computing resources (e.g., networks, servers, storage, applications, and services) that can be rapidly provisioned and released with minimal management effort or service provider interaction"}~\cite{NIST}.  Among its main characteristics, we have scalability and reliability; the first is granted by the possibility of creating multiple instances which can be easily distributed. 
Recently the cloud has been used as a platform abstracting underlying infrastructure resources, particularly for Serverless and Functions as a Service \cite{Serverless2020} \cite{Serverless2021}.

Fog computing is the computing layer between the cloud and the edge. The main goal of fog nodes is to minimize the load on the cloud by performing some services closer to the edge, providing a reduction both in streaming loads and response time.  

In Edge computing, the computation takes place at the edge of the network where the data is usually generated. Use the Edge is an effective solution whenever there are network problems and very strict response time. However, it is essential to highlight that edge computing does not have the computing and storing capabilities of the cloud or the fog. 
To solve the problems related to restrictions in timing, storage, or computational power the concepts of Cloud Continuum and Cognitive Cloud have been proposed.

Cloud Continuum has been defined as "\textit{an extension of the traditional Cloud towards multiple entities (e.g., Edge, Fog, IoT) that provide analysis, processing, storage, and data generation capabilities}" \cite{CCont}.

Cognitive Cloud, instead, is defined as "\textit{a Cloud-based system that is capable of sensing its environment, learning from it, and opportunistically and dynamically adapt its computational load as well as its outcome}" \cite{CognC}.

The main difference between these two definitions is that the Cloud Continuum is defined as the medium used to perform the computation while the Cognitive Cloud towards the capability of adapting the computational needs such as the load or the outcome.  

\subsection{Edge-to-cloud offloading}
\label{sec:offloading}

As mentioned in the previous section, Edge Computing extends computation facilities toward the edge of a network. Therefore, computation is performed near the end user, resulting in ultra-low latency and high bandwidth. Offloading algorithms allow end devices, edge nodes, and the cloud to work together. Generally, task offloading can be defined as the transfer of resource-intensive computational tasks to an external, resource-rich platform such as the ones used in Cloud, Edge, or Fog Computing.

There are different types of task offloading based on where the tasks are split and where is performed the computation; these are:
\begin{itemize}
    \item Partial offloading at the edge: In this type, part of the computation is executed locally at the end device, and remained will be offloaded at the edge.
    \item Full offloading at the edge: In this case, all computation tasks will be offloaded and executed at the edge.
    \item Partial/full offloading at the edge and at the cloud: Such offloading is for situations where the edge resources cannot execute all the tasks offloaded from the end device. Therefore, edge and cloud collaborate to process all the computational tasks \cite{saeik2021task}.
\end{itemize}
\subsubsection{Targets of task offloading}

There are different objectives for task offloading in the cloud continuum based on the different stakeholders, which can be categorized as follows \cite{saeik2021task}:
\begin{itemize}
    \item Delay: Minimizing the task execution delay is one of the main goals of task offloading. This delay can be split into different parts. It could be the delay related to the task execution at the device, the edge, or the cloud, delay related to the transmission at the various layers of the infrastructure, queuing delay, or task partitioning delay. The goal of reducing the delay by task offloading can be minimizing each of the mentioned delays or the average one \cite{yang2013framework}.
    \item Energy consumption: How to minimize the energy consumption by using task offloading is another meaningful objective of task offloading that typically refers to the end devices \cite{sardellitti2014distributed}. However, minimizing the energy consumption has to be followed by all the layers of this communication model because this problem is pushed to the edge and/or cloud infrastructure at the full offloading model \cite{mao2016dynamic, singh2013quality}.
   
    \item Bandwidth/spectrum: Allocation of spectrum in IoT and cellular networks plays an important role because of the limited bandwidth availability. Evaluating the spectrum utilization based on the number of offloaded tasks, power transmission, and bandwidth consumption is an efficient metric to deploy the available spectrum optimally \cite{sahni2017edge, zhao2017tasks, mao2017joint, zhang2017energy}.
\end{itemize}

\section{Research Questions}
\label{sec:ResearchQuestions}

Our goal is to identify edge-to-cloud tools available on the market and classify their characteristics. 

To achieve the aforementioned goal, we defined 3 main research questions (RQs).
\begin{itemize}
    \item [\textbf{RQ1.}] Which edge-to-cloud tools are available on the market? \\
    In this RQ, we aim at finding tools capable of performing offloading, orchestration, or other computational tasks along the Cognitive Cloud Continuum. The available tools are researched in the grey and peer-reviewed literature.
    
    \item [\textbf{RQ2.}] What are the characteristics of edge-to-cloud tools? \\
    In this RQ we aim at finding the characteristics of the tools which have been found from different sources. Whether they are just for offloading,  capable of orchestrating, or if they are simulators.
    
    \item [\textbf{RQ3.}]  What are the target environments of edge-to-cloud tools?\\
    In this RQ we aim at finding out the environment where the discovered tools can work.
 \end{itemize}
\section{Study Design}
\label{sec:StudyDesign}

This section outlines the process used for this work, which involved a Multivocal Literature Review (MLR) following the guidelines proposed by Garousi et al.~\cite{Garousi18a}, due to the topic's novelty. The following subsections provide a summary of the MLR process, including the selection of primary studies, quality assessment of the gray literature, data extraction, tool selection, conducting the review, and verifiability and replicability.

\subsection{MLR process overview}  
The MLR process encompasses a variety of sources, including peer-reviewed and gray literature, and acknowledges the diverse perspectives of both practitioners and academic researchers. To categorize contributions, MLR distinguishes between academic literature (peer-reviewed papers) and gray literature (other forms of content such as blog posts, white papers, podcasts, etc.).


\begin{figure}[ht]
    \centering
    \includegraphics[width=\columnwidth]{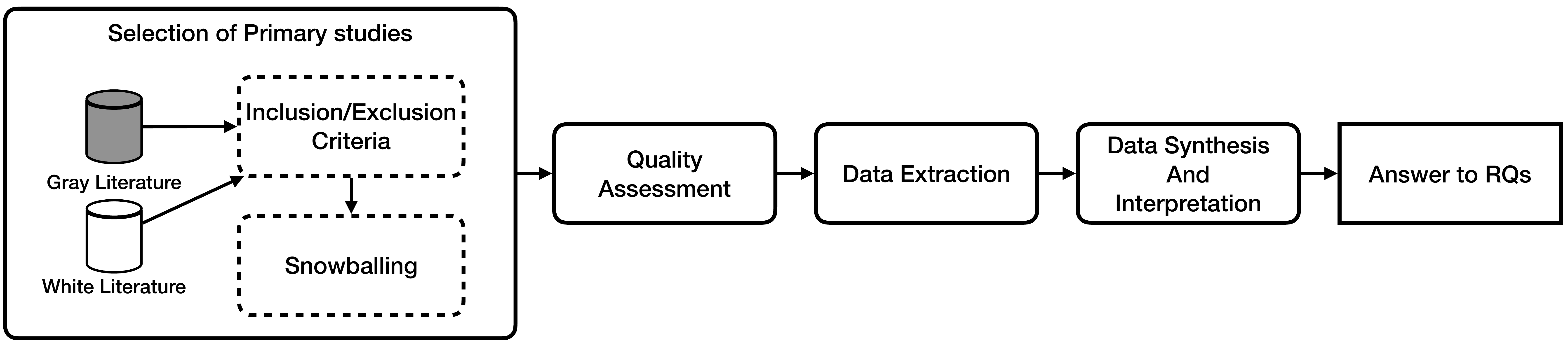}
    \caption{Overview of the followed MLR process}
    \label{fig:mlrprocess}
\end{figure}

\subsection{MLR motivation}
When it comes to exploring a complex and rapidly evolving field like Edge to Cloud Tools, it's crucial to approach the literature with a multivocal perspective. Multivocal literature reviews acknowledge diverse perspectives, opinions, and experiences of researchers and practitioners. By incorporating multiple voices, this approach can help to surface hidden assumptions, biases, and blind spots in the existing literature, as well as open up new avenues for inquiry and innovation.

In the context of edge-to-cloud tools, a multivocal literature review could of particular value. From the perspectives of software developers, hardware manufacturers, cloud providers, policymakers, end-users, and other stakeholders, the benefits, risks, and trade-offs of Edge to Cloud Tools can look very different. By engaging with multiple voices, this multivocal literature review can help to paint a more nuanced and comprehensive picture of Edge to Cloud Tools. It can reveal the divergent interests, values, and priorities that underpin different perspectives. This approach can also help to highlight gaps and contradictions in the existing literature and identify opportunities for further research and collaboration.

Ultimately, this work can help to enrich the understanding of Edge to Cloud Tools and inform more effective and inclusive approaches to their development and deployment. It can help us to embrace the complexity and diversity of this dynamic field.

\subsection{Selection of Primary Studies}
\label{sec:Selection}
The first step for selecting the Primary Studies (PS) is the search string identification that will be adopted in the academic bibliographic sources and in the gray literature source engines. 
We define the search string as follows:

\begin{center}
\textbf{("edge cloud" OR "edge-to-cloud") \\
AND \\
(offloading OR cognitive OR orchestration) \\
AND\\
tool}
\end{center}
To maximize the number of retrieved works, the search terms were used across all fields (i.e. title, abstract, and keywords). The same search terms were utilized for both gray literature from online sources and white literature from academic bibliographic sources, with both searches being performed in November 2022.

\textbf{Peer-reviewed literature search}. We considered the papers indexed by five bibliographic sources:

\begin{itemize}
    \item IEEEXplore digital library \cite{ieee}
    \item Scopus \cite{scopus}
    \item ACM digital library \cite{acm}
    \item Science Direct \cite{sciencedirect}
    \item ISI Web of science \cite{webofscience}
\end{itemize}

\textbf{Gray literature search}. 
We performed the search using two search engines: 

\begin{itemize}
    \item Google Search\footnote{\url{https://www.google.com/}} 
    \item Medium\footnote{\url{https://medium.com}}
\end{itemize}

The search results consisted of books, blog posts, forums, websites, videos, white-paper, frameworks, and podcasts.


\textbf{Application of inclusion and exclusion criteria}. Based on guidelines for Systematic Literature Reviews \cite{keele2007guidelines}, we defined \textit{inclusion and exclusion} criteria (Table \ref{tab:inclusionExclusion}). 
We considered less restrictive inclusion criteria to enable the inclusion of a more comprehensive set of tools.

To ensure the effectiveness of the inclusion and exclusion criteria, a subset of 10 randomly selected primary studies (PSs) from those retrieved were tested before their application. The resulting inclusion and exclusion criteria are outlined in Table~\ref{tab:coding}.

To screen each paper, two researchers were tasked with independently fairly reviewing them. The paper assignments were mixed up, and each researcher was given a similar number of papers to review, along with other team members. 
Cohen's kappa coefficient was calculated to assess inter-rater agreement. The results of this assessment can be found in the replication package~\footref{Package}.

\begin{table}[h]
\centering
\footnotesize
    \caption{Inclusion/Exclusion Criteria for Primary Studies Selection }
    \label{tab:coding}
    \begin{tabular}{l|p{8.4cm}}
    \hline 
\textbf{Primary Study (PS)} & \textbf{Criteria}\\ \hline
\multirow{1}{*}{\textbf{Inclusion}} & Research papers or search results that are commercial and open source tools performing computational tasks along the Cloud Continuum (Edge-to-Cloud) \\ \hline
\multirow{1}{*}{\textbf{Exclusion}} & Not in English  \\
 & Duplicated (post summarizing other websites)  \\
 & Out of topic (using the terms for other purposes) \\
 & Non peer-reviewed papers \\
 & Research Plans, roadmaps, vision papers\\
\hline

\end{tabular}
\label{tab:inclusionExclusion}
\end{table}

\subsection{Quality Assessment of the Gray Literature}
\label{sec:Quality}
In contrast to peer-reviewed literature, gray literature is not subjected to a formal review process, and its quality is less regulated. To assess the credibility and quality of selected gray literature sources and determine whether or not to include them, we followed the guidelines recommended by Garousi et al.~\cite{Garousi18a}, taking into account the authority of the producer, applied methodology, objectivity, date, novelty, and impact.

To evaluate each source, the first two authors utilized the aforementioned criteria, using either a binary or three-point Likert scale depending on the specific criteria. 

\subsection{Data Extraction}
As our goal is to characterize information from edge-to-cloud offloading tools, we need to get the information directly from the tools' websites. 
Therefore, the data extraction process is composed of two steps: 

\begin{itemize}
    \item[(PE)]\textit{Extraction of the list of tools from the primary studies} (PSs) that satisfied the quality assessment criteria. 
    \item[(TE)] \textit{Extraction of the information from the tools list}. In this case, we extracted the information directly from the official website portals.
\end{itemize}

We utilized a review spreadsheet to extract information following our research questions (RQs). A summary of the data extraction form can be found in Table~\ref{tab:ext}, as well as the mapping of the necessary information for addressing each RQ.

\begin{table}[]
  \footnotesize
    \centering
    \caption{Data Extraction }
    \label{tab:ext}
    \begin{tabular}{l|l|p{4.2cm}|l} \hline
\textbf{RQs}	& 	\textbf{Info}	&	\textbf{Description}	& \textbf{Step	}\\\hline
RQ$_1$ & Tool Name	&	Name of the tool & \multirow{2}{*}{PE}\\
RQ$_2$ & Tool Url &  \\\hline
RQ$_1$ & \textit{Where} to move & Categorizing the tools based on where they can move the computational tasks & \multirow{6}{*}{TE}\\ \cline{1-3}
RQ$_2$ & Characteristics & Identify the main characteristics of each tools & \\ \cline{1-3}
RQ$_3$ & Environment & Main environment for the tool & \\\hline

\end{tabular}
\end{table}

The data extraction process adhered to the qualitative analysis guidelines proposed by Wohlin et al.~\cite{wohlin2012experimentation}. All information was extracted by two researchers. If there was disagreement, a third author was consulted, and a discussion was held until the disagreement was resolved.

\subsection{Tool Selection}
To identify the final set of tools required to answer our RQs, a similar process to the one employed in the paper selection phase (Section~\ref{sec:Selection}) was applied, which involved filtering the tools based on a set of inclusion and exclusion criteria.

As with the PS selection process, we tested the applicability of the inclusion and exclusion criteria on a randomly selected subset of 10\% of the retrieved tools before applying them. The final set of inclusion and exclusion criteria is presented in Table \ref{tab:inclusionExclusion}.

The tool selection was carried out by two researchers who independently reviewed them, as was the case with the PS selection. Any discrepancies, a third author was brought in to reach a consensus. The inter-rater agreement was also assessed in this case by calculating Cohen's kappa coefficient. These results are documented in the replication package~\footref{Package}.

\subsection{Conducting the review}


From the Search process, conducted in November 2022, we retrieved a total of 1073 unique PS (after the exclusion of 137 duplicated): 220 PS from the white literature (Table~\ref{table:bibliographicResults}) and 853 PS from the gray literature (Table~\ref{table:2}).

Out of the 1073 works we retrieved and after applying inclusion and exclusion criteria, with an almost perfect agreement (Cohen's kappa = 0.765), the two authors agreed on excluding 771 works (157 from white literature and 614 from gray literature), resulting in \textbf{302} PS (63 from white literature, 239 from gray literature). 

From the data extraction process, 338 tools were obtained. 

The application of inclusion and exclusion criteria for tools resulted in a fair agreement (Cohen's kappa = 0.523) and a final set of 68 tools, as reported in Table~\ref{tab:license}.

\begin{table}[h] 
\centering
\caption{Initial search result from bibliographic sources.}
\begin{tabular}{lc}
\hline
\multicolumn{1}{c}{\textbf{Bibliographic Source}} & \textbf{\#non-duplicated papers} \\
\hline
IEEEXplore                             & 5                                        \\
Scopus                                 & 144                                      \\
ACM Digital Library                    & 69                                       \\
Science Direct                         & 1                                        \\
ISI Web of Science                     & 1                                       \\\hline
\textbf{Total}              & \textbf{220}                             \\\hline
\hline
\end{tabular}
\label{table:bibliographicResults}
\end{table}

\begin{table}[h]
\centering
\caption{Search results from Gray Literature Search Engines.}
\begin{tabular}{lc}
\hline
\multicolumn{1}{c}{\textbf{Search engines}} & \textbf{\#non-duplicated search result} \\
\hline
Google search                              & 495                                         \\
Medium search                              & 358                                   \\\hline
\textbf{Total}         & \textbf{853}                                             \\\hline
\hline
\end{tabular}
\label{table:2}
\end{table}


After applying the inclusion and exclusion criteria, we focused on one of the characteristics extracted during the previous stage: \textit{Tool Type}.
In the previous stage, we had 3 possible outcomes for the Tool Type characteristic, namely Commercial, Open Source, and Research Prototype. As a quality check, we included two more exclusion criteria as reported in Table~\ref{tab:quality}.
This would result in checking every single tool marked as Research Prototype and categorizing it as Commercial or Open Source. Whenever a tool would not respect any of the Quality Exclusion Criteria, it is automatically excluded. 

After the Quality Check, the final selected tools were 40.  

\begin{table}[h]
\centering
\footnotesize
    \caption{Inclusion/Exclusion Criteria for Tool Selection.}
    \label{tab:quality}
    \begin{tabular}{l|p{5.3cm}}
    \hline 
\textbf{Tools Selection} & \textbf{Criteria}\\ \hline
\multirow{1}{*}{\textbf{Inclusion}} & Commercial and Open Source Tools \\\hline
\multirow{1}{*}{\textbf{Exclusion}} & Not downloadable  \\
\multirow{1}{*}{\textbf{Exclusion}} & Open Source Tools with less than 100 stars in GitHub\\\hline

\hline

\end{tabular}
\label{tab:inclusionExclusion}
\end{table}

\subsection{Verifiability and Replicability}
\noindent To allow our study to be replicated, we have published the complete raw data in the replication package.\footnote{\url{https://figshare.com/articles/dataset/Replication_package/22567708}\label{Package}}.

\section{Study Results}
\label{sec:StudyResults}

\begin{table}[]
\caption{Tools characteristics}
\centering
\adjustbox{max width=\columnwidth}{
\begin{tabular}{|l||c|c||c|c|c|c|c|c|c|c|c|c||c|c|c|c||c|}
\toprule
& \multicolumn{2}{c||}{\textbf{RQ1}} & \multicolumn{10}{c||}{\textbf{RQ2}} & \multicolumn{4}{c||}{\textbf{RQ3}} & \\ \cline{2-17}
\textbf{		Tool Name	}	&	\textbf{	\rotatebox{90}{edge-to-cloud}	}  &	\textbf{	\rotatebox{90}{edge-only}	} &	\textbf{	\rotatebox{90}{Offloading}	}	&	\textbf{	\rotatebox{90}{Orchestration}	} &	\textbf{	\rotatebox{90}{Workflow Management}	}	&	\textbf{	\rotatebox{90}{Proprietary Container}}	&	\textbf{	\rotatebox{90}{Deployment}}	&	\textbf{	\rotatebox{90}{Kubernetes Distribution }}	&	\textbf{	\rotatebox{90}{Kubernetes Extension }}	&	\textbf{	\rotatebox{90}{Simulator}}	&	\textbf{	\rotatebox{90}{E2E Service}}	&	\textbf{	\rotatebox{90}{Platform}} & \textbf{	\rotatebox{90}{Agnostic}} & \textbf{	\rotatebox{90}{Cloud}}	& \textbf{	\rotatebox{90}{Edge Node}} & \textbf{	\rotatebox{90}{Far Edge}} & \textbf{	\rotatebox{90}{License}} \\	\midrule
\rowcolor{gray10}		Amazon EKS	&				&		x	&					&					&					&					&					&					&			x		&					&					&			&x & & & & Com		\\	
		Ambassador Edge Stack	&				&		x	&					&					&					&					&					&					&			x		&					&					&				&x & & & & Com	\\	
\rowcolor{gray10}		Apache Airflow	&			x	&			&					&					&			x		&					&					&					&					&					&					&			&x & & & & Com		\\	
		APEX	&			x	&	&					&					&					&					&					&					&					&					&			x		&				&x & & & & Com	\\	
\rowcolor{gray10}		Aruba ESP	&			x	&	&					&					&					&					&					&					&					&					&			x		&			& & x & & & Com		\\	
		Avassa	&				& x	&					&					&					&					&			x		&					&					&					&					&				&x & & & & Com	\\	
\rowcolor{gray10}		AWS IoT GreenGrass	&			x	&	&					&					&					&					&					&					&					&					&			x		&			& & & x & & Com		\\	
		AWS Wavelength	&			x	&	&					&					&					&					&					&					&					&					&			x		&			& & & x & & Com		\\	
\rowcolor{gray10}		Azure stack Edge	&			x	&	&					&					&					&					&					&					&					&					&			x		&		& & x & & & Com			\\	
		Baetyl		&				& x &					&					&					&					&					&					&			x		&					&					&			&x & & & & OSS		\\	
\rowcolor{gray10}		Cloudify	&			& x	&					&					&					&					&			x		&					&					&					&					&				&x & & & & Com	\\	
		Docker Swarm	&			x	&	&					&			x		&					&					&					&					&					&					&					&				&x & & & & Com	\\	
\rowcolor{gray10}		Eclipse ioFog 2.0	&			x	&	&					&			x		&					&					&			x		&					&					&					&					&			&x & & & & Com		\\	
		EdgeCloudSim	&			x	&	&					&					&					&					&					&					&					&			x		&					&				& &x & & & OSS	\\	
\rowcolor{gray10}		eKuiper	&			x	&	&					&					&			x		&					&					&					&					&					&					&				& & & & x & OSS	\\	
		FogFlow	&			x	&	&					&			x		&					&					&					&					&					&					&					&				&x & & & & OSS	\\	
\rowcolor{gray10}		Home edge orchesterator	&			x	&	&					&					&					&					&					&					&					&					&					&			x	&x & & & & OSS	\\	
		Intel Smart Edge	&			x	&	&					&					&					&					&					&					&					&					&					&			x	& & x & & & OSS	\\	
\rowcolor{gray10}		k0s	&			x	&	&					&			x		&					&					&					&			x		&					&					&					&			&x & & & & OSS		\\	
		K3s	&			x	&	&					&			x		&					&					&					&			x		&					&					&					&			&x & & & & OSS		\\	
\rowcolor{gray10}		KubeEdge	&			x	&	&					&			x		&					&					&					&			x		&					&					&					&			&x & & & & OSS		\\	
		KubeFed		&			x	& &					&		x			&					&					&					&			x		&					&					&					&				&x & & & & OSS	\\	
\rowcolor{gray10}		Kubernetes	&			x	&	&					&			x		&					&					&					&			x		&					&					&					&			&x & & & & OSS		\\	
		MicroK8S	&			x	&	&			x		&					&					&					&					&					&					&					&					&			&x & & & & OSS		\\	
\rowcolor{gray10}		Microsoft’s Azure IoT Edge	&			x	&	&					&					&					&					&					&					&					&					&			x		&				&x & & & & Com	\\	
		Nearby One	&			x	&	&					&			x		&					&					&					&					&					&					&					&			&x & & & & Com		\\	
\rowcolor{gray10}		Nomad	&			x	&	&					&			x		&					&			x		&					&					&					&					&					&			x	&x & & & & OSS	\\	
		NEBULA	&			x	&	&					&			x		&					&			x		&					&					&					&					&					&			x	&x & & & & OSS	\\	
\rowcolor{gray10}		Nuvlabox (NuvlaEdge)	&			x	&	&					&					&					&					&					&					&					&					&					&			x	&x & & & & Com	\\	
		ONAP	&			x	&	&					&			x		&					&					&					&					&					&					&					&				& & x & & & OSS	\\	
\rowcolor{gray10}		Open horizon	&			x	&	&					&					&					&					&					&					&					&					&					&			x	& & & x & & Com	\\	
		Open Stack Starlingx	&			x	&	&					&					&					&					&			x		&					&					&					&					&			& &x & & & OSS		\\	
\rowcolor{gray10}		OpenNebula	&			x	&	&					&					&					&					&					&					&					&					&					&			x	&x & & & & Com	\\	

		OpenShift	&			x	&	&					&					&					&					&					&					&					&					&					&			x	&x & & & & Com	\\	
\rowcolor{gray10}		Openstack	&			x	&	&					&					&					&					&					&					&					&					&					&			x	&x & & & & Com	\\	
		OpenYurt	&			x	&	&					&					&					&					&					&					&					&					&					&			x	&x & & & & OSS	\\	
\rowcolor{gray10}		Ormuco	&			& x	&					&					&					&					&					&					&					&					&			x		&			x	&x & & & & Com	\\	
		Saguna	&			x	&	&					&					&					&					&					&					&					&					&					&			x	& & & x & & Com	\\	
\rowcolor{gray10}		Windriver Studio	&			x	&	&					&					&					&					&					&					&					&					&					&			x	&x & & & & Com	\\	
		Zededa	&			x	&	&					&					&					&					&					&					&					&					&					&			x	&x & & & & Com	\\	
\bottomrule
\end{tabular}
}
\label{tab:char}
\end{table}

In this section, we present the results of our MLR study guided by the research questions stated in section \ref{sec:ResearchQuestions}. 
Table~\ref{tab:char} summarize the results achieved among the different RQs.  Such a Table can be subdivided into 4 main areas: the first one includes the information related to RQ1, the second is related to RQ2, the third to RQ3, and the last area is specifically reserved for the license.

\subsection{Edge-to-cloud tools (RQ1)}

\emph{Description}. To answer RQ1, we found out the tools that can perform offload, orchestration, or other computational tasks in the cloud continuum. 

\emph{Results}. The main objective of performing task offloading, orchestration, or workflow management is to optimize the computational tasks from an end-user device to a remote site under specific constraints. This process consists of three main parts that are (i) various hardware components, such as end-user devices and Edge/Cloud devices, (ii) multiple computing processes, including task splitting and computational processing either locally or remotely and (iii) networking components for transferring data between the hardware components involved \cite{saeik2021task}.

As discussed in section 2.2., the different types of computational tasks are categorized based on where they are executed. According to the final list of selected tools, we identify two types; {\em edge-to-cloud} and {\em edge-to-edge only}. As illustrated in Table~\ref{tab:char}, from the 40 tools, 6 of them are capable of moving the tasks only among edge devices, while 34 can move to the cloud.

\subsection{Tools characteristics (RQ2)}

\emph{Description}. To answer RQ2, we extracted the characteristics of the tools in the cloud continuum. 

\emph{Results}. We extracted characteristics of the identified tools and their alternatives, and grouped them into the following categories:
We identified 11 global categories:
\begin{itemize}

     \item \textbf{Offloading:} tools used to perform different kinds of offloading as introduced in Section~\ref{sec:offloading}.

    \item \textbf{Orchestration:} tools performing Orchestration. Tools that are capable of "managing, automating and coordinating the flow of resources between multiple types of devices, infrastructure, and network domains at the edge of a network" \cite{orch}.

    \item \textbf{Workflow Management:} tools used to manage the workflow and tasks assignable.
    \item \textbf{Proprietary Container:} tools that allow the creation of containerized applications.
    \item \textbf{Deployment:} tools used to manage and deploy containerized and non-containerized applications across the cloud continuum.

    \item \textbf{Kubernetes Distribution:} different distributions of Kubernetes designed to provide an out-of-the-box solution for deploying Kubernetes and managing containerized workloads.
    \item \textbf{Kubernetes Extension:} additional components that extend the core functionality of Kubernetes. These extensions are developed and added by third-party. Moreover, are designed to enhance the functionality of Kubernetes and provide additional capabilities for managing complex workloads.

    \item \textbf{Simulator:} tools used to simulate specific environments for testing. Such environments are useful when performing tests among different simulated entities and devices.
    \item \textbf{End2End (E2E) Service:} set of functionalities composed of multiple tools used to automate the full lifecycle of the devices along the cloud continuum. Such tools work together seamlessly to provide the desired functionality. The goal of an end-to-end service is to provide a complete solution to a specific problem or set of problems.
    \item \textbf{Platform:} complete software environment that provides a set of services and tools for developing, deploying, and managing distributed applications. The goal of a platform is to provide a high-level abstraction of the underlying infrastructure, making it easier for developers to build and deploy distributed applications.
    \item \textbf{License:} the license adopted by the different tools. A tool can be based on a Commercial (Com) or released as an Open Source Software (OSS).

\end{itemize}

From our extraction process, we discovered that only one tool is used to perform pure offloading: MicroK8s while on the contrary, 12 other tools are mostly used for orchestration purposes, and 2 for Workflow Management uses. On the same page, we discovered that the use of containers is of tremendous importance in this environment as 2 tools have been categorized as Proprietary Containers, and 4 tools are used to perform deployment but, most importantly, the high use of Kubernetes allowed us to label two specific categories: the Kubernetes distributions composed of 5 tools and the Kubernetes Extensions composed of 3 tools.
While only one tool can be categorized as a Simulator, multiple End-to-End Services (i.e. 7) and Platforms (i.e. 14) have been discovered, with Ormuco being categorized as both of the latter.
Out of the 40 retrieved tools, 23 have a commercial license, and 17 have an Open Source License. 
The mapping of each different tool to the different categories is depicted in the second part of Table~\ref{tab:char}. A complete list of tools and their URL is depicted in Table~\ref{tab:license}.

\subsection{Tools environment (RQ3)}

\emph{Description}. To answer RQ3, we identified the target environment for each tool. 

\emph{Results}. We categorized the different tools based on the main environment they target. We identified 4 different alternatives:
\begin{itemize}
    \item \textbf{Agnostic:} tools created to run seamlessly on different entities among the cloud continuum. We identified 29 agnostic tools.
    \item \textbf{Cloud Infrastructure:} tools that target the Cloud as the main environment to be run. Out of 40 final tools, 6 of them could be run at cloud infrastructures.
    \item \textbf{Edge Node:} tools made to run on Edge Devices. We found 5 tools whose main environment is edge nodes.
    \item \textbf{Far Edge:} tools targeting the deployment on those devices which reside in the Far Edge and therefore have lower computation capabilities. Based on our results, only 1 tool was made to run on the far edge.
\end{itemize}
Among the tools analyzed, the vast majority can be cataloged as Agnostic (29 out of 40), the following 6 tools have been labeled as targeted for the Cloud, 4 for the Edge Nodes, while only eKuiper targets a deployment in the Far Edge.

The second part of table~\ref{tab:char} maps each different tool to its targeted environment.
\section{Discussion}
\label{sec:Discussion}

The results achieved in this work provide insight view of the state of the market for what concerns the ability to move the computational burden of software among different parts of the Cloud Continuum. Given the hype that has targeted cloud computing for more than a decade, as expected, most tools target the execution of tasks on the cloud. However, given the strict requirement that nowadays are valid for edge and far-edge devices, this cannot always be a solution, and therefore the necessity of orchestrators and offloaders becomes essential.

It is therefore surprising to find only a single tool that targets as its primary task the offloading of tasks being executed, which is clearly a consequence of the inherent technical hurdles that are involved in such a process. While only 2 tools are able to perform as workflow managers, many more tools are capable of performing orchestration.
This shows that it is not always easy to dynamically adapt the computational tasks but is usually preferred to move the tasks before they actually start and plan in advance the execution and computation. Unsurprisingly a very important point arising from this MLR is the heavy presence of Kubernetes in the list of selected tools.
Kubernetes is, no coincidence, mostly known for its excellent orchestrator capabilities and all its distribution are of different use based on the necessity of the user or \textit{architect} of the environment together with these, many other tools extend or make use of Kubernetes' capabilities to perform valuable orchestration.

For what concerns container technologies, from our analysis we can see that all of the agnostic and cloud infrastructure-based technologies are compliant with the Open Container Initiative (OCI)~\cite{OCI} except for Nomad and NEBULA which have proprietary container structures. 

On the design of the tools, we can say that the difficulties of creating an End-to-end tool are reflected by the producer of the tool as those are mostly owned by very big companies. Ormuco is the only exception to this rule, being a company employing 34 people only (2022 data).


Regarding target platforms, platform-agnostic ones are the most common. This is reflected by the fact that the closer you move to hardware the more difficult it gets to support heterogeneous hardware/software platforms. This is shown also by only a few platforms targeting mainly edge nodes, and only one for the far edge.

\subsection{Feature challenges}

The analysis of the literature shows that the design of systems allowing for task offloading is still in its infancy. The tools are either targeted at a specific use case or are extensions of mainstream tools such as Kubernetes or containers. Moreover, the lack of interest in simulation hints that the scenarios under consideration are limited in size. 

To cover targeted use cases, most tools either provide coverage for end-to-end service or are full software platforms. This approach works against standardization and interoperability efforts, which would grow if it was possible to use different tools for the various actions related to task offloading and orchestration. 

Moreover, as stated in the previous section, the difficulties in creating a system capable of dynamically adapting the computational tasks (i.e. Cognitive), is nowadays reflected by the presence of multiple works researching how to perform advanced orchestration mechanisms aiming at achieving offloading but not so many tools released on the matter. We believe that a future challenge in the field will be related to the creation of tools leveraging AI-based algorithms for offloading purposes.

\subsection{Threats to validity}

We are aware that our work is subject to threats to validity, since we got through only tools that are available either commercially or as OSS, while other tools could be used internally by large companies, or more advanced techniques could be developed and close to reaching the maturity level required to be part of a tool such as the ones included in this analysis.

One more issue, that we tried to solve with a thorough analysis, was that some tools could have been abandoned, and slowly growing obsolescent. 

With regards to the tools' features, we based our analysis on the material available as a bibliography, and on the tools' website, not considering experimental features. This approach can be too defensive and produce several false negatives, nevertheless, we preferred this issue instead of accepting characteristics and capabilities not mature enough for tools that can be used in a production environment.

To improve the reliability of this work, we defined search terms and apply procedures that can be replicated by others. Since this is a mapping study and no systematic review, the inclusion/exclusion criteria are only related to whether the topic of Cognitive Cloud is present in a paper or not, as suggested by~\cite{petersen2008systematic}. 

As for the analysis procedure, since our analysis only uses descriptive statistics, the threats are minimal. However, we are aware that the synthesis of the definition might be subjective. To mitigate this threat, the analysis was done collaboratively, using a collecting coding method, and discussing with all the authors about inconsistencies. The Kohen K index about our disagreement also confirms the quality of the qualitative analysis performed.

\section{Conclusion}
\label{sec:Conclusion}
This paper presents the results of a Systematic Mapping Study to classify the edge-to-cloud tools in the cognitive cloud continuum. We conducted a Multivocal Literature Review (MLR) considering 40 tools from 1073 primary studies (220 PS from the white literature and 853 PS from the gray literature), as presented in section \ref{sec:StudyResults}. 

One of our main findings is that 85\% of the tools can perform offloading, orchestration, or other computational tasks from edge-to-cloud while the rest can execute on the edge only. We extended such a study to analyze the different characteristics of the selected tools to provide a valuable comparison for researchers and practitioners selecting the proper tool for their computational tasks. Such characteristics include the nature of the tools, but also the environment target for the deployment and its license.

\section*{CRediT authorship contribution statement}

\textbf{Sergio Moreschini:} Conceptualization, Methodology, Writing Original draft preparation.\\
\noindent \textbf{Elham Younesian:} Conceptualization, Methodology, Writing  Original draft preparation.\\
\textbf{David Hästbacka:} Conceptualization, Methodology, Writing  Original draft preparation.\\
\textbf{Michele Albano:} Conceptualization, Supervision Reviewing and Editing.\\
\textbf{Jiří Hošek:} Conceptualization, Supervision Reviewing and Editing. \\
\textbf{Davide Taibi:} Conceptualization, Supervision Reviewing and Editing.

\section*{Acknowledgement}
This work is part of project Industry X and 6GSoft, it is funded by Business Finland.

\bibliographystyle{model1-num-names}
\bibliography{main}

\section*{Appendix A: the Selected Papers}
{\small
\begin{enumerate}[labelindent=-5pt,label={[SP}{\arabic*]}]

\smallskip\item \label{SP1} 
Georgakopoulos, A. et al. (2012). Cognitive cloud-oriented wireless networks for the Future Internet. In 2012 WCNCW (pp. 431-435). IEEE.
\smallskip\item \label{SP2}
Cai, W. et al. (2014). Environment Perception for Cognitive Cloud Gaming. In International Conference on Cloud Computing (pp. 3-13). Springer, Cham.
\smallskip\item \label{SP3}
Cai, W. et al. (2014). Resource management for cognitive cloud gaming. In 2014 ICC (pp. 3456-3461). IEEE.
\smallskip\item \label{SP4}
Cordeschi, N. et al. (2015). Reliable adaptive resource management for cognitive cloud vehicular networks. IEEE Trans. on Vehicular Tech., 64(6), 2528-2537.
\smallskip\item \label{SP5}
Shi, W. (2015). QoE guarantee scheme based on cooperative cognitive cloud and opportunistic weight particle swarm. Electrical and Computer Engineering, 2015.
\smallskip\item \label{SP6}
Baughman, A. K. et al. (2015). Disruptive innovation: Large scale multimedia data mining. In Multimedia Data Mining and Analytics (pp. 3-28). Springer, Cham.
\smallskip\item \label{SP7}
Mahmoodi, S. E. et al. (2016). A time-adaptive heuristic for cognitive cloud offloading in multi-RAT enabled wireless devices. IEEE Transactions on Cognitive Communications and Networking, 2(2), 194-207.
\smallskip\item \label{SP8}
Chiang, M., et al. (2016). Fog and IoT: An overview of research opportunities. IEEE Internet of things journal, 3(6), 854-864.
\smallskip\item \label{SP9}
Mahmoodi, S. E. et al. (2018). Cognitive Cloud Offloading Using Multiple Radios. In Spectrum-Aware Mobile Computing (pp. 23-33). Springer, Cham.
\smallskip\item \label{SP10}
Wu, X. et al. (2018). Phase-compensation-based cooperative spectrum sensing algorithm for cognitive cloud networks. In 2018 ICOIN (pp. 755-759). IEEE.
\smallskip\item \label{SP11}
Wang, L. et al. (2018). Cooperative Spectrum Sensing Algorithm Based on Phase Compensation in Cognitive Cloud Networks. ICUFN (pp. 143-147).
\smallskip\item \label{SP12}
Huang, H. et al. (2018). On the Performance of Cognitive Cloud Radio Access Networks in the Presence of Hardware Impairment. In APSIPA ASC (427-431).
\smallskip\item \label{SP13}
Marshall, T. E. et al. (2018). Cloud-based intelligent accounting applications: accounting task automation using IBM watson cognitive computing. Journal of Emerging Technologies in Accounting, 15(1), 199-215.
\smallskip\item \label{SP14}
Jann, J. et al. (2018). IBM POWER9 system software. IBM Journal of Research and Development, 62(4/5), 6-1.
\smallskip\item \label{SP15}
Kloeckner, K. et al. (2018). Building a cognitive platform for the managed IT services lifecycle. IBM Journal of Research and Development, 62(1), 8-1.
\smallskip\item \label{SP16}
Mahmoodi, S. E. et al. (2019). Time-Adaptive and Cognitive Cloud Offloading Using Multiple Radios. In Spectrum-Aware Mobile Computing (pp. 49-66).
\smallskip\item \label{SP17}
Garai, Á. et al. (2019). Revolutionizing healthcare with IoT and cognitive, cloud-based telemedicine. Acta Polytechnica Hungarica, 16(2), 163-181.
\smallskip\item \label{SP18}
Amato, F. et al. (2019). A federation of cognitive cloud services for trusting data sources. In CISIS (pp. 1022-1031).
\smallskip\item \label{SP19}
Ferrer, A. J. et al. (2021). Towards a Cognitive Compute Continuum: An Architecture for Ad-Hoc Self-Managed Swarms. In CCGrid (pp. 634-641).
\smallskip\item \label{SP20}
Campolo, C. et al. (2021). Virtualizing AI at the distributed edge towards intelligent IoT applications. Journal of Sensor and Actuator Networks, 10(1), 13.
\smallskip\item \label{SP21}
Vermesan, O. et al. (2021). Internet of Vehicles–System of Systems Distributed Intelligence for Mobility Applications. In Intelligent Technologies for Internet of Vehicles (pp. 93-147). Springer, Cham.
\smallskip\item \label{SP22}
Kretsis, A. et al. (2021). SERRANO: Transparent Application Deployment in a Secure, Accelerated and Cognitive Cloud Continuum. In MeditCom (pp. 55-60)
\smallskip\item \label{SP23}
Bacciu, D. et al. (2021). TEACHING-Trustworthy autonomous cyber-physical applications through human-centred intelligence. In 2021 COINS (pp. 1-6).
\smallskip\item \label{SP24}
Zhang, H. et al. (2021). Knowledge-based systems for blockchain-based cognitive cloud computing model for security purposes. International Journal of Modeling, Simulation, and Scientific Computing, 2241002.

\end{enumerate}
}
\newpage
\section*{Appendix B: Tools list}
\begin{table*}[!ht]
\caption{Tools main reference and their license}
\label{tab:license}
\centering

\adjustbox{width=\textwidth}{
\begin{tabular}{|l|l|}
\toprule		
Tool                & URL   \\
\toprule

Amazon EKS                                          & {\url{https://aws.amazon.com/eks/features/}}                                                                                                                       \\

Ambassador Edge Stack                                                       & {\url{https://www.getambassador.io}}                                                                                                                               \\

Apache Airflow                                                              &{\url{ https://airflow.apache.org}}                                                                                       \\

APEX (for Dell Technologies) & {\url{https://www.dell.com/en-us/dt/apex/}}                                                                                                   \\
Aruba Edge Services Platform (ESP)                                          & {\url{ https://www.arubanetworks.com/solutions/aruba-esp/}}                                                                                                   \\
Avassa                                                                      &{\url{ https://avassa.io}}                                                                                                                                     \\
AWS IoT GreenGrass                                                          & {\url{ https://aws.amazon.com/greengrass/}}                                                                                                                                            \\
AWS Wavelength                                                              &{\url{ https://aws.amazon.com/wavelength/}}                                                                                                                                             \\
Azure stack Edge                                    &{\url{ https://azure.microsoft.com/en-us/products/azure-stack/edge}}                                                                                                  \\
Baetyl                                                                      & {\url{ https://baetyl.io/en/}}                                                                                                                                     \\
Cloudify                                            & {\url{ https://cloudify.co/}}                                                                                                                                                       \\
Docker Swarm & {\url{ https://docs.docker.com/engine/swarm/}}                                                                                                                         \\
Eclipse ioFog 2.0                                                           & {\url{https://iofog.org}}                                                                                              \\
EdgeCloudSim                                                                &{\url{https://github.com/CagataySonmez/EdgeCloudSim}}                                                           \\
eKuiper                                                                     & {\url{ https://ekuiper.org/}}                                                                                                 \\
FogFlow                                                                     &{\url{https://github.com/smartfog/fogflow}}                                                                                    \\
Home edge orchesterator                                                     & {\url{ https://www.lfedge.org/projects/homeedge/}}                                                                            \\
Intel Smart Edge                                                            &{\url{ https://smart-edge-open.github.io}}                                                                                     \\
k0s                                                                         &{\url{ https://k0sproject.io/}}                                                                                                \\
K3s                                                                         & {\url{ https://k3s.io/}}                                                                                                      \\
KubeEdge                                                                    &{\url{ https://kubeedge.io/en/}}                                                                                               \\
KubeFed                                                                     &{\url{ https://github.com/kubernetes-sigs/kubefed}}                                                                            \\
Kubernetes                                                                  &{\url{ https://kubernetes.io/}}                                                                                                \\
MicroK8S                                                                    &{\url{ https://microk8s.io/}}                                                                                                  \\
Microsoft’s Azure IoT Edge                                                  &{\url{ https://azure.microsoft.com/en-us/products/iot-edge}}                                                            \\
Nearby One                                                                  &{\url{ https://www.nearbycomputing.com/nearbyone/}}                                                                     \\
NEBULA                                              &{\url{ https://nebula-orchestrator.github.io/}}                                                                                                        \\
Nomad                                                                       &{\url{ https://www.nomadproject.io/}}                                                                                          \\
Nuvlabox (NuvlaEdge)  &{\url{https://nuvla.io/ui/sign-in}}                                                                                                                                           \\
ONAP (Open Network Automation Platform)             &{\url{ https://www.onap.org/}}                                                                                                                         \\
Open horizon                                                                & {\url{ https://www.lfedge.org/projects/openhorizon/}}    \\
Open Stack Starlingx                                & {\url{ https://www.starlingx.io/}}                                                                                                                                     \\
OpenNebula Edgify                                                           & {\url{ https://opennebula.io/edgify-opennebula-as-a-service/}}                                                                                                          \\
OpenShift                                           & {\url{ https://www.redhat.com/en/technologies/cloud-computing/openshift}}                                                                                                                       \\
Openstack                                           & {\url{ https://www.openstack.org/}}                                                                                                                                                             \\
OpenYurt                                            & {\url{ https://openyurt.io/}}                                                                                                                                                                          \\
Ormuco                                                                      &{\url{ https://ormuco.com}}                                                                                                      \\
Saguna                                                                      &  \url{https://www.saguna.net/product/saguna-edge-to-cloud/}                                                                                   \\
StudioGA (WINDRIVER STUDIO)                                                 & {\url{ https://www.windriver.com/studio}}                                                                                                     \\
Zededa                                                                      & {\url{ https://zededa.com/}}                                                                                                                 \\
\bottomrule
\end{tabular}}
\end{table*}

\end{document}